\documentclass[journal]{IEEEtran}
\usepackage{amsmath,amsfonts}
\usepackage{algorithmic}
\usepackage{algorithm}
\usepackage{array}
\usepackage[caption=false,font=footnotesize,labelfont=sf,textfont=sf]{subfig}
\usepackage{textcomp}
\usepackage{stfloats}
\usepackage{xurl}
\usepackage{verbatim}
\usepackage{graphicx}
\usepackage{cite}

\usepackage{xcolor}
\usepackage{MnSymbol}
\usepackage{tikz}
 
\begin{document}

\title{Fast Network Recovery from Large-Scale Disasters: A Resilient and Self-Organizing RAN Framework}
\author{M. Yaser {Yagan},~\IEEEmembership{Graduate~Student~Member,~IEEE}, Sefa Kayraklik,~\IEEEmembership{Graduate~Student~Member,~IEEE},\\ Samed Kesir,~\IEEEmembership{Graduate~Student~Member,~IEEE}, Gizem Sumen,~\IEEEmembership{Graduate~Student~Member,~IEEE}, \\Ibrahim {Hokelek},~\IEEEmembership{Member,~IEEE}, Mehmet {Basaran}, George C. Alexandropoulos,~\IEEEmembership{Senior~Member,~IEEE}, \\{Ertugrul} {Basar},~\IEEEmembership{Fellow,~IEEE}, Cicek Cavdar,~\IEEEmembership{Member,~IEEE}, Huseyin Arslan,~\IEEEmembership{Fellow,~IEEE}, \\and Ali Gorcin, ~\IEEEmembership{Senior~Member,~IEEE}
\thanks{M. Y. Yagan, S. Kayraklik, S. Kesir, G. Sumen, I. Hokelek and A. Gorcin are with the Communications and Signal Processing Research (HİSAR) Lab., TÜBİTAK-BİLGEM, Kocaeli, Türkiye. (e-mail: \{yaser.yagan, sefa.kayraklik, samed.kesir, gizem.sumen, ibrahim.hokelek, ali.gorcin\}@tubitak.gov.tr). A. Gorcin is also with the Department of Electronics and Telecommunications Engineering, Istanbul Technical University, Istanbul, Turkey.}
\thanks{M. Basaran is with the 6GEN Laboratory, Next-Generation R\&D, Network Technologies, 34854 Istanbul, Türkiye (e-mail: mehmet.basaran@turkcell.com.tr).}
\thanks{G. C. Alexandropoulos is with the Department of Informatics and Telecommunications, National and Kapodistrian University of Athens, 15784 Athens, Greece and with the Department of Electrical and Computer Engineering, University of Illinois Chicago, IL 60601, USA (e-mail: alexandg@di.uoa.gr).}
\thanks{E. Basar is with the Communications Research and Innovation Laboratory (CoreLab), Department of Electrical and Electronics Engineering, Koç University, Sariyer, Istanbul, Türkiye. (e-mail: ebasar@ku.edu.tr).}
\thanks{C. Cavdar is with the Department of Computer Science, KTH Royal Institute of Technology, Kista, Sweden (e-mail:cavdar@kth.se).}
\thanks{H. Arslan is with the Department of Electrical and Electronics Engineering, Istanbul Medipol University, 34810 Istanbul, Türkiye (e-mail: huseyinarslan@medipol.edu.tr).}
\vspace{-24pt}
}

\maketitle

\begin{abstract}
Extreme natural phenomena are occurring more frequently
everyday in the world, challenging, among others, the infrastructure of communication networks. For instance, the devastating earthquakes in Turkiye in early 2023 showcased that, although communications became an imminent priority, existing mobile communication systems fell short with the operational requirements of harsh disaster environments. In this article, we present a novel framework for robust, resilient, adaptive, and open source sixth generation (6G) radio access networks (Open6GRAN) that can provide uninterrupted communication services in the face of natural disasters and other disruptions. Advanced 6G technologies, such as reconfigurable intelligent surfaces (RISs), cell-free multiple-input-multiple-output, and joint communications and sensing with increasingly heterogeneous deployment, consisting of terrestrial and non-terrestrial nodes, are robustly integrated. We advocate that a key enabler to develop service and management orchestration with fast recovery capabilities will rely on an artificial-intelligence-based radio access network (RAN) controller. To support the emergency use case spanning a larger area, the integration of aerial and space segments with the terrestrial network promises a rapid and reliable response in the case of any disaster. A proof-of-concept that rapidly reconfigures an RIS for performance enhancement under an emergency scenario is presented and discussed.
\end{abstract}


\section{Introduction}

\IEEEPARstart{N} \protect atural disasters around the world, such as earthquakes, floods and fires, also affected by climate change, manifest the limitations of existing mobile communication systems and underline the importance of network survivability, resilience and recovery time, which also constitute key requirements on the agenda of the upcoming sixth generation (6G) networks. %Natural disasters and extreme weather events can cause damages to communication infrastructure resulting in severe disruptions in telecommunication services. 
Furthermore, a massive demand for wireless communication services from/to regions under state of calamity typically overloads communication networks, causing interference, signal attenuation, delays, link failures, and breaks. To this end, new network survivability metrics and models are needed as a measure of the 6G, and beyond, network resilience against failures due to large-scale disaster events.

The recent devastating earthquakes in Turkiye in early $2023$ hit its southeast region, directly affecting $11$ large cities that comprise a sixth of the country’s total population \cite{TurkiyeReport}. According to the network traffic reports from the database of
Turkcell, the leading mobile network operator of Turkiye, there was an explosive increase in the peak rates of data and voice traffic, reaching 260\% and 9150\%, respectively, following the first half an hour of the earthquake compared to the previous day. 
% REVISED - TURKCELL
As a result of the substantial collapse of buildings in three primary urban areas (Adiyaman, Hatay, and Kahramanmaras), overall data traffic activity notably decreased compared to the corresponding day of the previous week. Likewise, following the earthquake, there was a slight decrease in 3G and 4G LTE signal availability for the same cities compared to the previous week. However, 2G availability experienced a more pronounced decline due to the denser deployment of 2G-compliant base stations (BSs). Furthermore, this circumstance notably impacted LTE throughput and diminished call success rates. On the first day of the earthquake, the proportion of customers unable to receive any cellular signal was markedly elevated, particularly in the three major cities affected.

Communication performance and coverage declines resulted from collapsed buildings that housed the small BSs on top of them. The operational BSs that remained until the earthquake could only function for about four hours using their internal energy reserves, due to the post-earthquake electricity shortage, provided they had not crashed due to the high volume of voice and data requests from users. It is noteworthy that gas and generator supplies quickly became insufficient, as they could not be replenished due to closed highways hindering access to any relevant aid. An attempt was made for the first time to integrate a Turkcell civilian BS into a medium-altitude drone named Aksungur, developed by Turkish Aerospace Industries, Inc. (TUSAS) \cite{TUSAS23}.

%The performance and coverage losses stem from the collapsed buildings hosting the small base stations (BSs) on top of them. Whereas the remaining operational BSs were able to only serve around four hours with their own internal energy supply due to the electricity shortage after the earthquake if they were not crashed because of the intense voice and data requests of the users. It is important to note that gas and generator supplies became insufficient quickly since they could not be replaced by spare or alternative ones due to the closed highways obstructing the aid from accessing. The integration of a Turkcell civilian base station into a medium-altitude drone, was attempted for the first time. However, interference issues with surviving terrestrial base stations were encountered indicating that optimizing the drone's trajectory is required to provide a reliable connectivity.

%Natural disasters around the world, like earthquakes, floods, and fires, affected also by climate change, manifest the limitations of the existing mobile communication systems and underline the importance of network survivability, resilience, and recovery time, which also constitute key requirements on the agenda of upcoming 6G networks. 
As a natural hazard renders a wireless network non-operational, completely or partially, self-monitoring and self-healing strategies as well as the inclusion of new network nodes on the fly are necessary towards creating a robust, resilient, sustainable, and elastic network infrastructure. The challenges here arise from various emergency use cases and scenarios, such as simultaneous multinode failures, interference issues, unprecedented traffic patterns, as well as sensing and positioning needs in addition to electricity shortages and logistic issues. 6G technologies, such as non-terrestrial networks (NTNs)~\cite{6G_NTN}, cell-free multiple-input-multiple-output (cfMIMO)~\cite{elhoushy2021cell}, reconfigurable intelligent surfaces (RISs)~\cite{basar2023RIS_mag}, and joint communications and sensing (JCAS)~\cite{6g-disac} provide new opportunities towards creating a self-organizing and sustainable wireless network with reliable connectivity, high data rates, localization, and sensing. For example, NTN enables 3D connectivity with the capability of adding new nodes on the fly for providing larger coverage area in addition to a rapid formation of backhaul links. A fast reconfiguration of the cfMIMO and large mobile RISs will be instrumental in providing ubiquitous and robust connectivity under an operationally difficult environment where sensors or robots can be deployed to identify life or transfer rescue supplies (e.g. food / water). The on-the-fly emergency network will provide communications among people and rescue teams, but can also enable JCAS to transfer multi-modal sensing data from sensors and devices thrown in the field to synthesize the tracking of people, goods, etc. In such circumstances, the use of millimeter wave (mmWave) frequencies is inevitable, despite their challenges, to support the higher data rate requirements of emergency communications for both access and backhaul.
\begin{figure}[!t]
\centering
\includegraphics[width=0.85\linewidth]{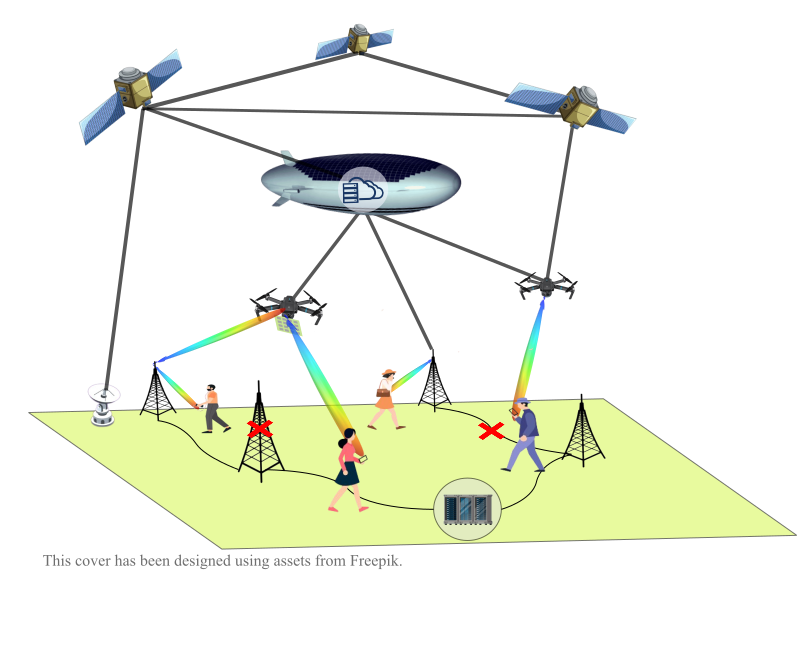}
\caption{A fast recovery aerially integrated network in a post-disaster scenario.}
\label{fig_5}
\end{figure}

In this article, we discuss the key ingredients of a robust, resilient, self-organizing, and open 6G radio access network (Open6GRAN), capable of surviving large-scale disasters by seamlessly integrating aerial and space network segments with mobile terrestrial network infrastructure. When the terrestrial communication network fails due to an emergency use case, as depicted in Fig. \ref{fig_5}, the first task is to identify the impact of the disaster event on the network infrastructure, thus determining the operational and non-operational BS and the out-of-service area. Several NTN nodes can be deployed in the disaster area to autonomously form a backhaul network using mmWave links, which can provide connectivity to the core network. Since higher frequencies suffer from severe path attenuation and are highly sensitive to blockage effects, mobile RISs with low-cost passive reflecting elements can be leveraged to create alternative line-of-sight (LOS) links and overcome severe signal blockage effects and path attenuation \cite{basar2023RIS_mag}. In traditional cellular communication systems, extreme traffic surges in disaster zones typically cause spectrum scarcity and inter-cell interference (ICI) due to overlapping coverage areas. Therefore, in order to provide seamless connectivity and improve spectrum efficiency with high reliability, cfMIMO systems can be utilized that combine multiple access points (APs) with massive antennas to a single central processing unit (CPU) without cell boundaries~\cite{elhoushy2021cell}. %Therefore, the CPU-based centralized control enables all APs to serve all UEs over the same time-frequency resources for more sustainable communications and spectrum efficiency. 
JCAS systems that effectively integrate radio communications and radar detection in a single system to further improve spectrum efficiency can also contribute to reducing the impact of hazardous phenomena and facilitate the access of critical information to emergency responders before and after a disaster \cite{cui2021integrating}. Toward this end, effective JCAS systems that offer high-resolution imaging as well as precise localization, detection, and positioning are needed to facilitate relief coordination and access to survivors and available resources. Most importantly, the latter 6G technologies need to be autonomously orchestrated via artificial intelligence (AI) and/or machine learning (ML)to form an emergency-respond network without the human in the loop, and reconfigure it rapidly as the network conditions change. However, current approaches typically concentrate on individual 6G RAN technologies and concepts separately, and a cross-layer design is missing. Considering the interaction of multiple technological elements with each other, their harmonization by a cloud-based O-RAN service management and orchestration mechanism~\cite{ORAN} is crucial for enabling diverse 6G use cases, optimizing network performance, and providing network resiliency under large-scale disaster events.  

%Artificial intelligence (AI), machine learning (ML), cloud and edge networking, virtualization, and Open RAN (O-RAN) will be other 6G enabling concepts from the system perspective to autonomously form a 6G emergency network without the human in the loop, and reconfigure it rapidly as the network conditions change. Nevertheless, the existing approaches typically concentrate on individual 6G RAN technologies and concepts separately while a cross-layer design is usually omitted. Considering the interaction of multiple technology elements with each other and their harmonization by a cloud-based O-RAN service management and orchestration mechanism is crucial for enabling diverse 6G use cases, optimizing network performance, and providing network resiliency under large-scale disaster events. 

\section{The Proposed Open6GRAN Network}
\begin{figure*}[!t]
\centering
\includegraphics[width=0.75\linewidth]{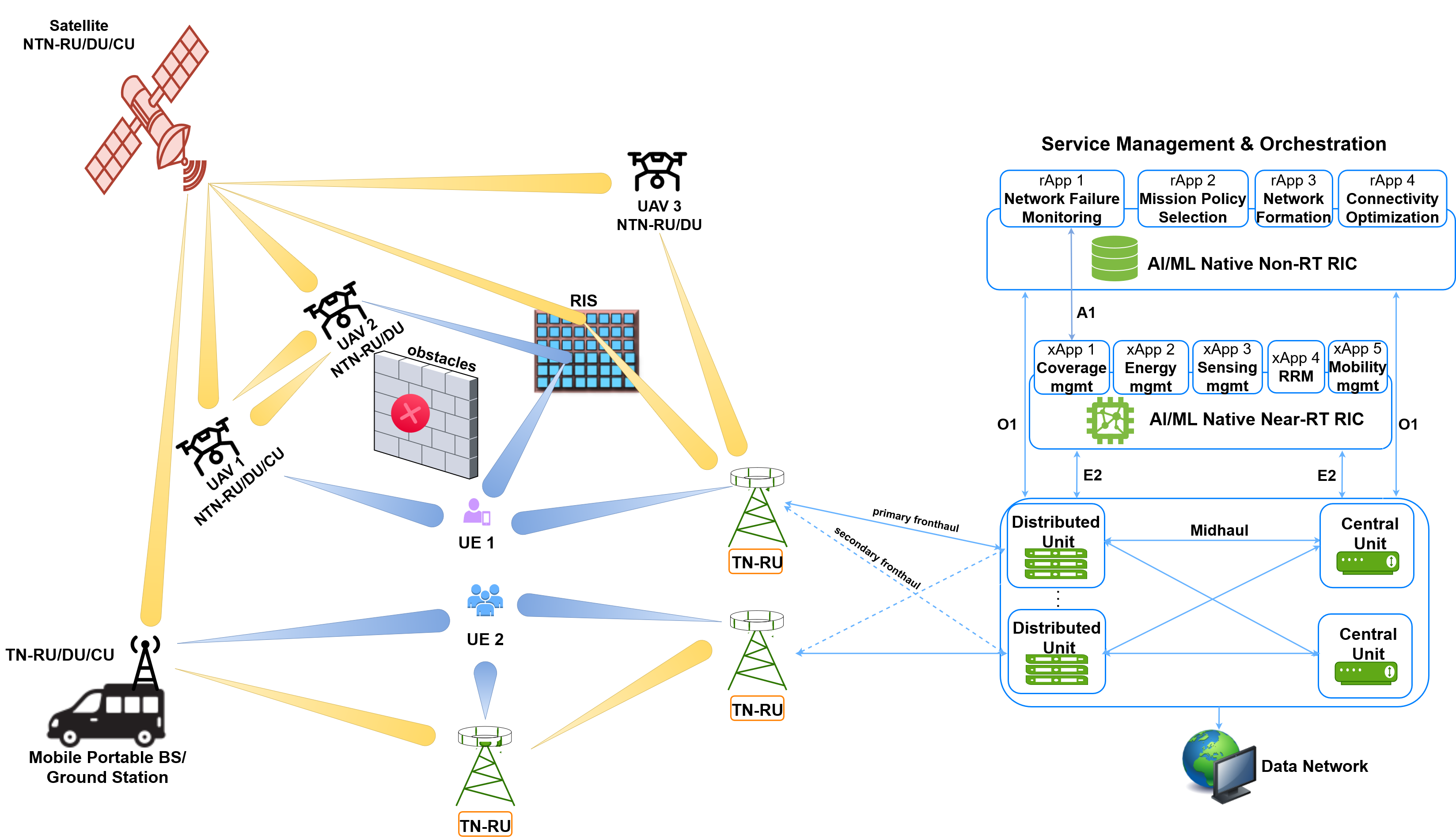}
\caption{The proposed Open6GRAN networking framework incorporating mobile portable BSs, UAVs, multi-fumctional RISs, and satellites.}
\label{fig_6}
\end{figure*}
The end-to-end architecture of the proposed robust,
resilient, and self-organizing Open6GRAN is illustrated in Fig.~\ref{fig_6}, where the cloud RAN (C-RAN) environment hosts disaggregated network components including distributed units (DUs) and central units (CUs), while radio units (RUs) are distributed over various terrestrial nodes (TNs) and NTN nodes. Mobile portable BSs, unmanned aerial vehicles (UAVs), and satellite nodes can potentially include DU and CU in addition to RU (e.g., TN-RU/DU/CU for a mobile portable BS and NTN-RU/DU/CU for UAV and satellite). This distributed MIMO approach will allow efficient utilization of diverse resources in the space, time, and frequency domains by combining the functionalities of the physical layer with advanced antenna concepts and capabilities. The main open problems in the realm of how to implement the existing physical-layer methods under practical circumstances, such as under heterogeneous fronthaul/backhaul limitations (capacity and delay), computational capabilities that are divided over many cloud servers, and challenging user services that require end-to-end optimization of the network resources, need to be addressed. This is a particularly important challenge, as we ambitiously aim to integrate TNs and NTN noes using distributed RUs.

The communication among the disaggregated RAN components (i.e., between RU and DU, DU and CU, CU and RAN intelligent controllers (RIC)) will be provided through both open interfaces defined by O-RAN Alliance and the F1 interface (a.k.a. mid haul interface) for DU-CU communication by 3GPP~\cite{ORAN}. All TNs and NTN RUs can be utilized to form a backhaul communication network, while only TNs and UAVs can be used for access network. RIS is an important enabler forming both backhaul and access links to overcome non-LOS issues. Each user equipment (UE) can be simultaneously served by multiple RUs using cfMIMO technology. In this architecture, RIC is another important concept in the service management and orchestration (SMO), where the AI-based management of limited radio resources can be performed using third-party applications, such as xApps for near-real-time RIC and rApps for non-real-time RIC. UAVs and satellites as part of NTN will be deployed to serve UEs for seamless communication continuity as an integrated access and backhaul network~\cite{FD_IAB} which is optimized using non-real-time (Non-RT) RIC applications, including network failure monitoring, mission policy selection, network formation, and connectivity optimization in addition to near-real-time (Near-RT) RIC applications, including radio resource, coverage, energy, sensing, and mobility management. The NTN support will be provided through novel waveforms designed based on distributed JCAS techniques to dynamically balance the trade-offs between communication efficiency and sensing accuracy through cooperative sensing with distributed RUs and AI/ML-based RIC applications~\cite{6g-disac}. When disaster occurs, urgent actions will be highly needed to continue the network services which are crucial especially for people's lives. Instead of direct and manual external attempts to build up the collapsed terrestrial BSs, which is indeed not easy due to challenging transportation issues as Türkiye experienced the recent giant earthquake, self-organization of the networks becomes a critical factor. When some of the terrestrial BSs go down, their position needs to be determined by benefitting from the JCAS techniques. Through intelligently managing RAN by xApps/rApps, sensing of channel state information is important for NTN overlays with related neighbor interference estimation. Then, it will be possible to transfer the exploding network traffic to aerial mobile BSs curring mmWave massive MIMO and multi-functional RISs~\cite{basar2023RIS_mag}, such as UAVs and satellites.

\section{O-RAN with Advanced 6G Technologies}
The technologies of RIS, cfMIMO, and JCAS are expected to play an important role in meeting the key performance and value indicators expected for 6G, including network survivability, resilience, and recovery time in large-scale disasters. The current state-of-the-art mostly concentrates on utilizing a single technology to support the envisioned use cases. Instead of the ``one-size-fits-all'' paradigm, the Open6GRAN framework proposes an open and standardized SMO mechanism which can harmonize multiple technologies through intelligent software-based agents. An O-RAN-based intelligent orchestration will enable fast and autonomous reconfiguration of 6G network resources as the network conditions change dynamically due to failures, mobility, and unprecedented traffic increases. AI/ML algorithms will be instrumental to provide efficient resource allocation under time-varying network conditions, where the algorithms can learn and generate predictions through real-time data collection and analysis defined in the O-RAN framework. 

As an example description of the Open6GRAN resource orchestration framework, there are two different time granularities which can be mapped to real-time and non-real- time RIC terminologies in the O-RAN concept. First, the O-RAN-based Non-RT rApps will perform network planning and optimization tasks for determining operational and non-operational TNs as well as the number of required NTN nodes for the emergency network recovery. After the NTN nodes are deployed over the disaster area, the O-RAN-based real-time xApps will perform fine granular cfMIMO and RIS resource optimization for providing quality of service (QoS) support in the emergency network. 

\subsection{Emergency Network Planning and Optimization with NTN}
Integration of terrestrial and non-terrestrial systems using mobile BSs is another big challenge that needs to be tackled to provide a sustainable 6G communication infrastructure~\cite{6G_NTN}. This integration will be instrumental in supporting emergency use cases that span a larger geographic area. During the recent devastating earthquake in Turkiye, although there were mobile BSs equipped on terrestrial vehicles, it took a significant amount of time to deploy those systems and additional efforts were needed to address the challenges related to the coexistence of aerial and terrestrial segments. The key takeaway from that tragedy was that NTN is required to provide a sustainable communication infrastructure in the event of disaster. 

In conventional integrated aerial/terrestrial systems, the addition of nodes (usually aerial) is done when the propagation / coverage / capacity is well provisioned~\cite{6G_NTN}. Even then, it is a challenging task to identify the optimum placement of the additional node and its transmission parameters including power, antenna tilt/direction, frequency reuse etc. However, in a disaster scenario, the challenge is further exacerbated, since the physical propagation environment is drastically different due to the collapse of buildings, trees, and even the ground terrain, which implies that the shadowing and multipath characteristics of the environment are no longer known~\cite{hong2016path}. Moreover, the network infrastructure might be damaged (or antennas misaligned in the case of wireless backhaul), which means that the network coverage is also vastly different. In this case, the aerial BSs firstly need to acquire information regarding the environment, either using radio measurement or 3D imaging to form a coverage map. This requires the development of on-the-fly sophisticated learning and data fusion techniques. Following this, the optimization of the aerial BS locations, trajectory, and transmission parameters needs to be carried out. 

Infrastructure damage and communication outages are a real possibility in cases of natural disasters, which requires the deployment of alternatives that are more robust, flexible, mobile, and deployable efficiently to complement the remaining infrastructure. This is where the O-RAN with aerial and space nodes comes into the picture, where the satellites can provide large coverage footprints and are immune to any natural disaster, while UAVs can be swiftly deployed. Fortunately, the O-RAN architecture relieves the network operators of any restriction in terms of the vendor which can be critical in a time-sensitive scenario such as those of natural disasters. 

%Fig. \ref{fig_18} illustrates how the ground, air, and space segments can be used in conjunction with each other.

In our proposed Open6GRAN framework, rApps will be utilized for network planning and optimization activities. Although aerial platforms provide speedy and flexible deployment capabilities, their integration requires overcoming significant challenges with regard to interference management and deployment optimization.
%it is still important to ensure that they are deployed optimally.
For instance, in the recent earthquake in Turkiye, even though aerial BSs were present, their integration into the network was a major challenge due to the interference theyr experienced from the survived terrestrial BSs. Therefore, ensuring the coexistence of the existing network with newly deployed nodes, especially in time-sensitive scenarios, remains a challenge. In general, there may be a requirement to provide both backhaul and fronthaul connectivity using aerial or spaceborne platforms in the aftermath of a disaster. The former can be realized using geostationary earth orbit (GEO) satellites and high-altitude platform stations (HAPSs), while low-earth orbit (LEO) satellites and UAVs are more suited for fronthaul connectivity. UAVs are the critical component of this framework because they are always autonomous, efficient, and connected to the network. 

%\begin{figure}[!t]
%\centering
%\includegraphics[width=0.95\linewidth]{fig18o6g.png}
%\caption{Illustration of how the different ground, air and space segments of the network can be integrated.}
%\label{fig_18}
%\end{figure}

\subsection{Fast Reconfiguration of Radio Resources}
%6G advanced technologies such as RIS, cfMIMO, and JCAS are the main building blocks towards meeting 6G KPIs and KVIs such as Tbps throughput, sub-ms latency, extremely high reliability, survivability, and short recovery time. 
RISs, which can exploit the wireless propagation environment by collectively adjusting the phases of incident signals with low-cost and low power consumption elements, are considered to be an important enabler for potential 6G use cases. They can preferably be used as standalone support modules (plug-and-play) to assist communication within the network without requiring significant signalling overhead~\cite{RIS_CC}. A promising way to achieve this goal is by adjusting the RIS phase shifts using a predetermined and finite set of reflection patterns (RPs), forming a codebook. The technology of cfMIMO utilizes a vast number of distributed antennas, which are spread out across a wide area, to provide high-quality wireless communication services~\cite{elhoushy2021cell}. A UE is served by multiple spatially distributed APs simultaneously. One of the key benefits of cfMIMO is to enable the use of advanced beamforming techniques, which can direct radio waves in specific directions to improve signal strength and reduce interference. This is particularly important in areas with large UE densities or where there are many obstructions, such as in urban environments. Another advantage is to provide seamless connectivity, as a UE can switch from one distributed antenna group to another without interruption. This leads to higher coverage and enhanced quality of experience, particularly in areas with varying signal strengths or where there coverage holes. The cognitive/intelligent networks of the future will heavily rely on awareness of the propagation environment that can be extracted from the wireless transmissions themselves via sensing. While the coexistence and cooperation are considered, the joint design of sensing and communication is arguably the most popular and interesting approach from a physical-layer design perspective, where the same waveform is used to perform both functionalities \cite{memisoglu2023waveform}. The Open6GRAN framework proposes to utilize O-RAN based SMO to seamlessly integrate RISs and cfMIMO into the 6G communication stack. RISs with low-cost passive reflecting elements can be used to overcome severe path loss and blockage effects at high frequencies, while cfMIMO utilizing multiple APs and massive antennas without cell boundaries can provide seamless connectivity, ultra-reliability, and enhanced spectrum efficiency. The AI/ML-based xApps will play a key role for the optimum usage of available radio resources by enabling fully cooperative networking along with the resource elasticity.
%\begin{figure}[!t]
%\centering
%\includegraphics[width=0.95\linewidth]{fig17o6g.png}
%\caption{The O-RAN based orchestration of RIS and cfMIMO.}
%\label{fig_17}
%\end{figure}

In a more challenging use case, radio resources need to be reconfigured according to time-varying network conditions due to mobility, variations in traffic demands, and failures. For example, the beamforming weights in cfMIMO and the phase configuration of the RIS elements need to be updated in a timely manner as the vehicles and UEs move in the network. The substantial changes in the network may require switching from RIS to cfMIMO, or vice versa. Open6GRAN, using the AI/ML-based radio resource controller framework, will provide timely reconfiguration of radio resources to achieve near-optimum performance. In the extreme case of a disaster scenario, a new UAV-based communication infrastructure will be deployed in the disaster area and integrated with the existing operational terrestrial network. In this use case, radio resources will be reconfigured on the fly to provide coverage and capacity optimization. In this framework, an emergency network will be created, not only to provide communications among people and rescue teams, but also enable JCAS for transferring multi-modal sensing data from sensors and devices thrown in the field to keep track of people, goods, floods etc.
\vspace{-6pt}
\section{RIS-Enabled QoS Enhancement Results}
In this section, preliminary measurement results demonstrating how a sub-6GHz RIS prototype can be rapidly reconfigured to provide QoS enhancement for emergency networks are presented and discussed. %Alternatively, these enhancements can also be provided through the fast re-configuration of the cfMIMO whose prototype is currently under development.

\vspace{-8pt}
\subsection{Fast RIS Optimization for Mobile UEs}
The phase configuration optimization of the RIS elements is an important research avenue, in which iterative, codebook-based, and AI/ML-based algorithms can be employed \cite{Alexandropoulos2022Pervasive,sefaRIS,kesir2023rapid}. During the training phase, an iterative approach utilizes the reference signal received power of the receiver as feedback to measure the fitness of the RIS configuration in each iteration. The grouping iterative algorithm creates logical groups whose elements are configured to provide the same phase shift before reflecting the incoming signals. On the other hand, in codebook-based schemes, offline RIS configurations, which are computed and stored for the selected locations, are rapidly applied for mobile receivers without requiring any feedback. The qualitative comparison of the RIS phase shift optimization algorithms is shown in Table~\ref{tab:comparisonRIS}. As shown, the iterative algorithm requires a longer training time and a higher signaling overhead to obtain a near-optimal RIS configuration. The grouping iterative algorithm reduces the number of iterations, and hence, the training time, however, the performance is slightly degraded compared to the iterative algorithm. The codebook-based algorithm is faster and completely eliminates or minimizes signaling overhead, while the performance quality is limited with the number of codebooks and the accuracy of receiver locations~\cite{RIS_CC}. The AI/ML approach, which is trained offline, is a potential alternative which can balance the trade-off among the performance, recovery (reconfiguration) time, and signaling overhead.

\begin{figure}[t]
\centering
\subfloat[]{\label{fig:RIS_codebook:a} 
\includegraphics[width=0.85\linewidth]{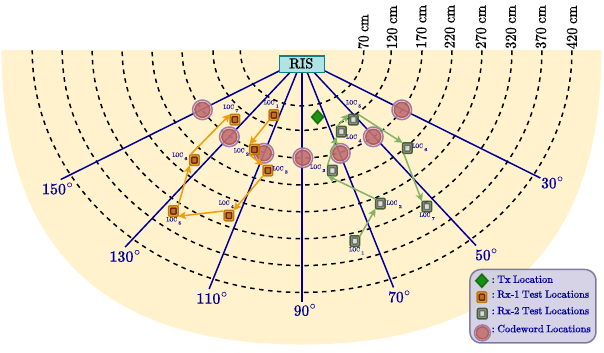}}\\
\subfloat[]{\label{fig:RIS_codebook:b} 
\includegraphics[width=0.85\linewidth]{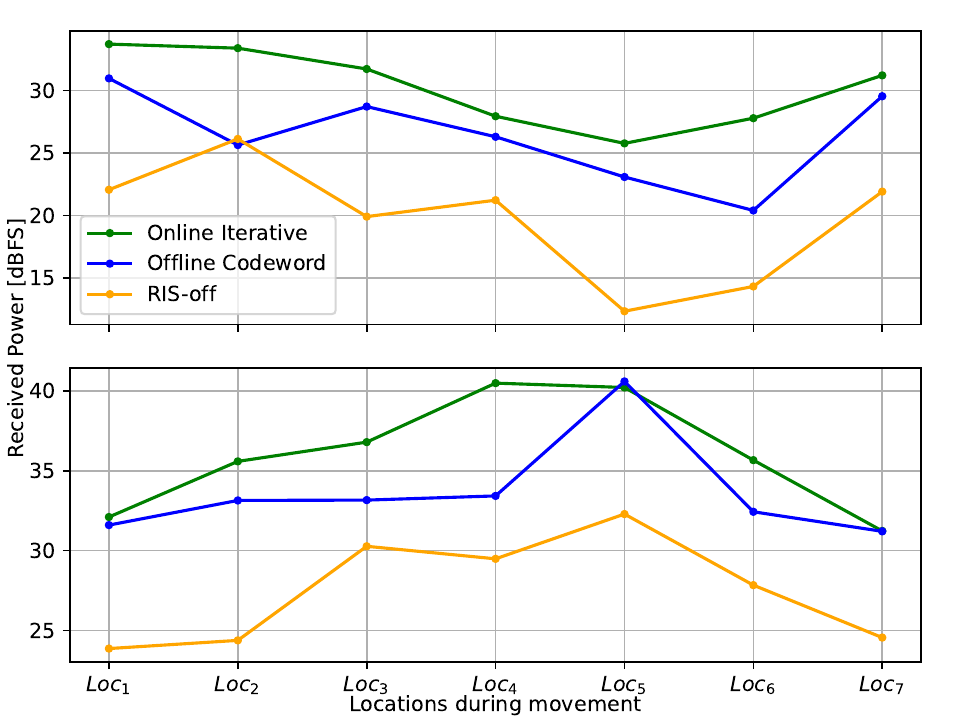}}
\caption{(a) The experimental scenario along with the trajectories of the two UEs (Rx1 and Rx2); \cite{sefaRIS} (b) the power measurements of the received signal of Rx1 (top) and Rx2 (bottom).}
\label{fig:RIS_codebook}
\vspace{-12pt}
\end{figure}

 % will be edited
The codebook-based RIS configuration is demonstrated for an RIS-aided multi-user wireless communication system. The RIS is partitioned into two parts for serving two UEs, namely Rx1 and Rx2. For the Rx1, a codebook including the codeword configurations corresponding to seven different grid angles at the distance of $170$ cm is generated using the first half of the RIS to boost the received signal power of Rx1, while the second codebook is formed using the other half to boost the received signal power of Rx2. 
%Five corresponding codewords for $38$ elements of the RIS are recorded for each location specified above since the same performance of the codewords cannot be achieved when the codewords for Rx1 and Rx2 are combined and applied to the RIS. Therefore, the best combination of the codewords of the Rx1 and Rx2 are selected from the five codewords of the Rx1 and Rx2. 
For the measurement experiments, Rx1 and Rx2 moved along the trajectories, as depicted in Fig.~\ref{fig:RIS_codebook}(a). During the movement of the Rx1 and Rx2, the codewords corresponding to the closest reference points were applied to increase their received signal powers. The measurement experiments were repeated for three cases: \textit{i}) the RIS is in no-phase-shift mode; \textit{ii}) the RIS is configured using the pre-computed codewords of the codebook; and \textit{iii}) the RIS is configured using the online iterative method on the test point. Figure\ref{fig:RIS_codebook}(b) shows the received signal powers of the Rx1 and Rx2 at the top and bottom plots, respectively, through the seven test points. Evidently, the codebook approach can successfully boost the received signal powers of both UEs. The iterative method provides a lot more received power compared to the offline codeword approach, as it takes significantly longer to calculate the corresponding RIS configurations. However, the codebook-based approach is more appropriate for emergency communication scenarios as it takes only milliseconds to reconfigure the RIS using the pre-computed codewords. The JCAS technology in the Open6GRAN framework can be instrumental in providing and maintaining an accurate location of mobile UEs for the codebook-based RIS configuration. When the location information of the UEs is unknown, searching the limited number of codebooks can take considerably shorter time than the iterative algorithm. The RIC module of the proposed Open6GRAN framework can be effectively designed to switch among the alternative RIS configuration algorithms if time-varying performance objectives are to be utilized (e.g., fast reconfiguration versus higher throughput).\vspace{-12pt}

%The offline codeword method for the Rx1 performs slightly better than the online iterative method since the iterative method behaves equally to the Rx1 and Rx2 through their joint optimization; therefore, the improvement obtained from the Rx2 being pushed into the noise floor is much higher than the Rx1. }
\begin{table}[t!]
\caption{RIS Phase Shift Optimization Algorithms.}
    \centering
    \begin{tabular}{c|c|c|c}
         \textbf{Algorithms} & \textbf{Performance} & \textbf{Recovery Time} & \textbf{Overhead}  \\
         \hline
         Iterative \cite{sefaRIS} & \normalsize $\filledstar \filledstar \filledstar\, \filledstar$ & \normalsize $\filledstar \smallstar \smallstar\, \smallstar$ & \normalsize $\filledstar \smallstar \smallstar\, \smallstar$ \\
         \hline
         Grouping Iterative \cite{sefaRIS} & \normalsize $\filledstar \filledstar \smallstar\, \smallstar$ & \normalsize $\filledstar \filledstar \smallstar\, \smallstar$ & \normalsize $\filledstar \filledstar \smallstar\, \smallstar$ \\
         \hline
         Codebook-based \cite{sefaRIS} & \normalsize $\filledstar \smallstar \smallstar\, \smallstar$ & \normalsize $\filledstar \filledstar \filledstar\, \filledstar$ & \normalsize $\filledstar \filledstar \filledstar\, \filledstar$ \\
         \hline
         AI/ML \cite{kesir2023rapid} & \normalsize $\filledstar \filledstar \filledstar\, \smallstar$ & \normalsize $\filledstar \filledstar \filledstar\, \smallstar$ & \normalsize $\filledstar \filledstar \filledstar\, \smallstar$ 
    \end{tabular}
    \label{tab:comparisonRIS}
    \vspace{-8pt}
\end{table}

\subsection{RIS-Empowered Wireless Communications}
%\begin{figure}[t]
%\centering
%\subfloat[]{\label{fig:RIS_srsRAN:a} 
%\includegraphics[width=0.85\linewidth]{H00_Const_RIS_Off.png}}\\
%\subfloat[]{\label{fig:RIS_srsRAN:b} 
%\includegraphics[width=0.85\linewidth]{H00_Const_RIS_On.png}}
%\caption{The channel response of the received signal and the constalletion diagram of the equalized symbols when the RIS is (a) off and (b) on.}
%\label{fig:RIS_srsRAN}
%\end{figure}

%\subsection{Throughput Demonstration of the RIS-assisted Wireless Communication System}

\begin{figure}[t]
\centering
\includegraphics[width=\linewidth]{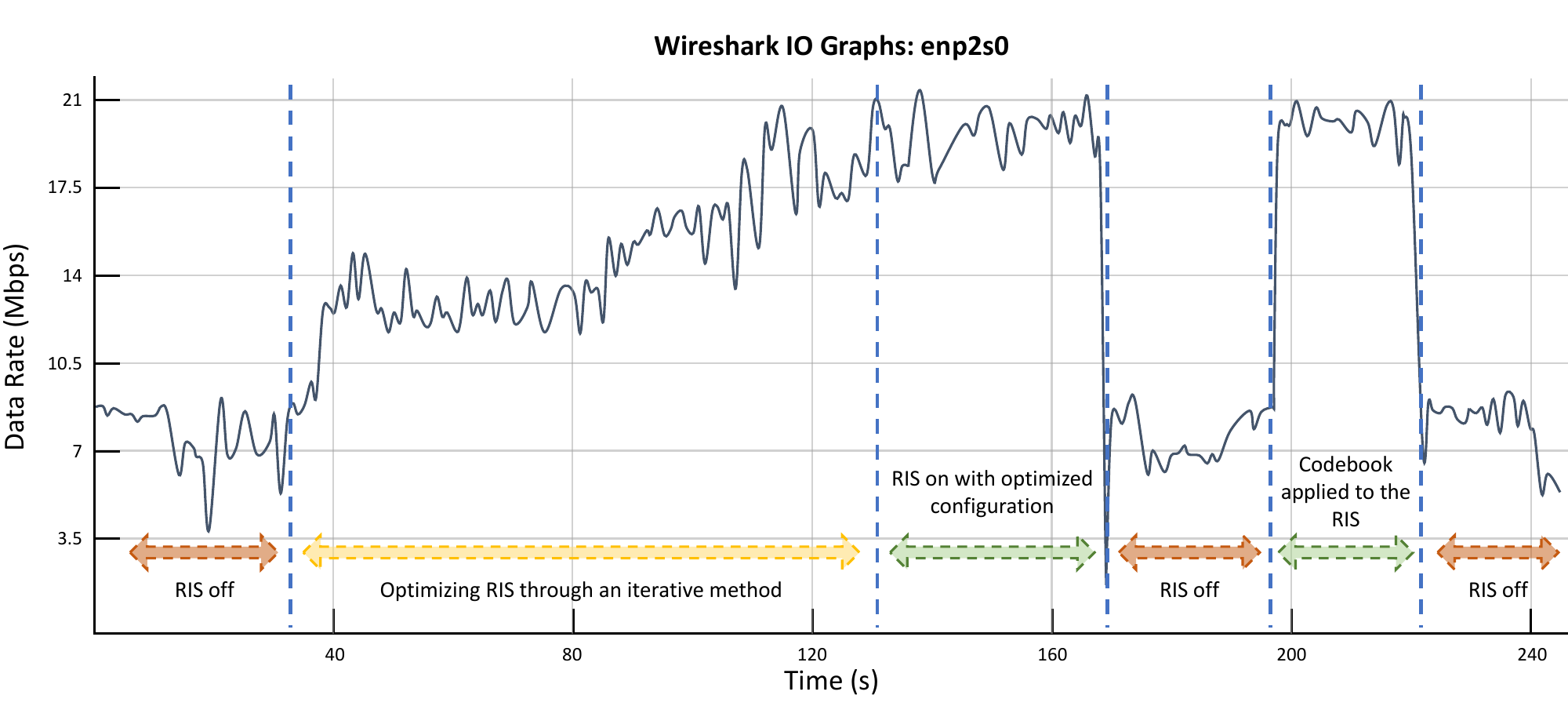}
\caption{Throughput analysis using the Wireshark application for the considered RIS-assisted wireless communication system.}
\label{fig:throughputRIS}
\end{figure}

Wireless transceivers supporting adaptive modulation and coding were utilized with and without the sub-6GHz RIS prototype in an indoor environment setting for a real-time demonstration. Figure~\ref{fig:throughputRIS} shows the real-time data rate measured at the receiver using the Wireshark program, where initially the data rate was lower as the RIS was deactivated. When the iterative algorithm was used to explore different RIS configurations, the data rate started increasing gradually, as the algorithm measured the received signal power at each iteration and kept the best RIS configuration providing the highest power. It took about $100$ seconds to try $76$ elements, each of which had four different configurations corresponding to horizontal and vertical polarizations in addition to phase shifts of $0$ and $180$ degrees for each polarization. Once the final RIS configuration was reached, the data rate varied between $18$ and $21$ Mbps due to the characteristics of the wireless channel that vary with time. When the RIS was turned off, the data rate immediately dropped below $10$ Mbps. When the RIS was turned on and its configuration was optimized via the iterative algorithm, the data rate suddenly jumped at $20$ Mbps showcasing that the codebook-based  approach can potentially provide a fast recovery for emergency communication scenarios. This result is particularly important as high resolution video services from/to regions under state of calamity may only be supported through the RIS assistance. %We have been working on extending this preliminary demonstration to realize the fast RIS reconfiguration through the xApps which communicates with the BS emulator through the E2 interface. 

\vspace{-6pt}
\section{Conclusion}
In this article, we have presented the Open6GRAN networking framework to create self-organizing and survivable wireless networks with fast recovery capabilities for emergency communication after large-scale disaster events. Inspired by real-life challenges faced by mobile operators, the proposed framework integrates TNs and NTN nodes for network planning and optimization and highlights how advanced 6G technologies, including RIS, cfMIMO, and JCAS can be used for fast recovery of networks from large-scale disasters. 

\vspace{-6pt}
\section*{Acknowledgments}
The authors would like to thank the researchers Dr. Ahmet Faruk Coskun, Dr. Zehra Yigit, and Dr. Fatih Cogen for their constructive support during this work from 6GEN LAB, Turkcell.
\vspace{-12pt}

\bibliographystyle{IEEEtran}
\bibliography{refs}

\end{document}